\documentclass{article}
\usepackage{spconf,amsmath,amssymb,graphicx,booktabs,url,tikz}
\usepackage{colortbl}
\usepackage{tabularx}
\usetikzlibrary{arrows.meta,positioning}
\usepackage[T1]{fontenc}
\usepackage{newtxtext,newtxmath}
\usepackage[hidelinks]{hyperref}
\usepackage{comment}

\usepackage{hyperref}
\hypersetup{
    colorlinks=false,
    pdfborder={0 0 1},
    citebordercolor={0 0.6 0},
    linkbordercolor={0 0.6 0}
}

\title{Tracing Decoder Artifacts for Compact Synthetic Speech Screening}
\name{Yi Chen Liu, Jian Liu}
\address{University of Georgia, Athens, GA, USA}
\begin{document}
\ninept
\maketitle
\begin{abstract}
Recent advances in speech synthesis and voice cloning have increased the need for reliable synthetic-speech detection, yet high-accuracy detectors increasingly rely on large pretrained models that are costly to invoke on every recording. Rather than replacing such detectors, we investigate a compact front-end screen that processes all inputs cheaply and forwards only suspicious recordings for more expensive analysis. To enable lightweight screening without a large learned encoder, we exploit spectral traces introduced by speech-generation operations. We analyze how learned upsampling and inverse short-time Fourier transform synthesis can produce predictable spectral artifacts and measure their presence directly in generated waveforms. Because the strength of these artifacts varies across generators, we combine decoder-guided spectral measurements with complementary descriptors of short-time spectral shape and temporal variation in a compact gradient-boosted tree. Across seven speech generators and two human-speech sources, the proposed screen achieves an equal error rate of 0.021\% with an estimated model storage of 151 KiB. When used as the first stage of a simulated cascade with a 1.15-billion-parameter detector, it reduces estimated detection energy by 84.4\% while operating at a 0.050\% synthetic-speech miss rate, demonstrating the potential of decoder-guided acoustic evidence for low-cost front-end screening.
%Advances in speech synthesis and voice cloning create opportunities for impersonation and fraud, making synthetic speech detection important for trustworthy voice communication. Although large speech detectors achieve high accuracy, running them on every recording can be energy and cost intensive. We propose a lightweight screening model that uses traces left by speech generators to distinguish generated speech from human speech. By examining speech decoders, we identify how these traces can arise and measure their presence in the generated audio. We combine these measurements with acoustic characteristics and their changes over time to build a compact detector. On an evaluation covering seven speech generators and two human speech sources, our detector achieves an in-domain equal error rate of 0.021\%. Its estimated model storage is 151 KiB, roughly one-fifth of the storage required by the smallest of four neural speech detectors evaluated in this study. In a simulated scenario, screening recordings before forwarding suspected synthetic speech to a 1.15-billion-parameter detector saves an estimated 84\% of detection energy compared with using the larger detector alone.
\end{abstract}
\begin{keywords}
%speech deepfake detection, decoder artifacts, spectro-temporal analysis, lightweight screening, handcrafted features
synthetic speech detection, decoder artifacts, lightweight screening, spectro-temporal analysis
\end{keywords}

\vspace{-3mm}
\section{Introduction}
\label{sec:intro}
\vspace{-2mm}
Recent advances in text-to-speech synthesis and voice cloning have enabled
high-quality AI voices for applications such as voice assistants, customer
service, online meetings, and social media. At the same time, increasingly
realistic synthetic speech creates new opportunities for impersonation and
fraud, making reliable synthetic-speech detection important for trustworthy
voice communication. High-performing detectors increasingly rely on large
pretrained speech encoders to capture rich acoustic representations. For
example, Speech DF Arena, a recent benchmark that evaluates open-source and
proprietary speech deepfake detectors across diverse datasets and attack
conditions~\cite{dowerah2026speech}, shows that many leading systems contain
hundreds of millions to billions of parameters. Although such models provide
strong detection performance, applying them to every incoming recording can
impose substantial computational and energy costs, particularly in
high-volume or always-on applications.

Rather than replacing these high-capacity detectors, we investigate a different deployment strategy: using a compact model as a front-end screen. The screen processes every recording locally and forwards only suspicious inputs to a more expensive backend detector. Such a design can reduce unnecessary large-model inference and, when screening is performed locally, can also reduce the amount of audio that must be transmitted for remote analysis. Importantly, the objective of such a screen differs from that of a standalone detector. A useful front end should remain inexpensive enough to run on every input while maintaining a sufficiently low synthetic-speech miss rate, so that most benign recordings can bypass expensive downstream processing.

This requirement raises a fundamental representation question: \textit{what acoustic evidence can support reliable screening without relying on a large learned encoder?} We explore traces introduced by the speech-generation process itself. Prior work shows that vocoder-resynthesized speech can provide useful evidence for detecting speech generated by complete text-to-speech and voice-conversion systems~\cite{wang2023spoofed}. In particular, waveform decoders commonly use learned upsampling or inverse short-time Fourier transform (iSTFT) synthesis, whose periodic operations can leave structured spectral traces in generated audio~\cite{pons2021upsampling}. These traces are attractive for lightweight screening because they can be measured directly from the waveform and linked to identifiable operations in the synthesis pipeline.

Decoder artifacts alone, however, provide incomplete evidence. Their frequencies and prominence depend on decoder architecture, sampling rates, learned filters, and subsequent waveform processing. We therefore combine mechanism-guided artifact measurements with complementary acoustic descriptors. Specifically, we analyze decoder upsampling and iSTFT synthesis to identify frequencies at which narrow spectral peaks may occur, and measure both their local prominence and persistent spectral residuals in the output waveform~\cite{afchar2025fourier}. We further characterize short-time spectral shape using linear-frequency cepstral coefficients (LFCCs) and summarize their temporal changes through local statistics and modulation spectra, following prior work on temporal variation in synthetic-speech detection~\cite{wu2013synthetic}. These complementary measurements capture both mechanism-specific spectral traces and broader spectro-temporal differences, and are fused using a compact gradient-boosted tree.

We evaluate the resulting screen on speech from seven generators and two human speech sources. It achieves an equal error rate of 0.021\% with an estimated model storage of 151~KiB, approximately 78\% smaller than the smallest neural comparator evaluated in our study. We further examine generalization to unseen generators and analyze the model as the first stage of a detection cascade. In a simulated deployment with a 1.15-billion-parameter backend detector, operating the screen at a 0.050\% synthetic-speech miss rate reduces estimated detection energy by 84.4\%. Together, these results show that decoder-guided acoustic measurements can provide a compact first-stage screening mechanism that avoids unnecessary large-model inference while retaining interpretable connections to the speech generation process.

\vspace{-3mm}
\section{Related Work}
\label{sec:related}
\vspace{-2mm}
Recent speech deepfake datasets broaden coverage to more speakers,
diverse acoustic conditions, and adversarial
attacks~\cite{wang2024asvspoof}. Large-scale multilingual collections
also cover text-to-speech, voice conversion, and neural vocoder outputs,
enabling evaluation across languages and synthesis
methods~\cite{huang2025speechfake}.
Evaluations across multiple datasets reveal substantial performance
variation across sources and recording conditions, showing that strong
results on one benchmark do not ensure cross-domain
reliability~\cite{dowerah2026speech}.

Self-supervised learning based detectors use speech encoders to
provide representations for high-accuracy deepfake
detection~\cite{el2025comprehensive}.
These encoders learn from large unlabeled speech corpora, and their
representations are used by classifiers trained on labeled detection data. Lightweight detectors pursue lower computational and storage costs
through compact networks or explicit acoustic features. Waveform
architectures use spectro-temporal graph attention or lightweight
convolutions~\cite{jung2022aasist,xiao2025rawtfnet}, while recent
cepstral studies examine MFCCs, LFCCs, and temporal differences for
seen-source and cross-source detection~\cite{schafer2025mfcc}.
Closely related artifact-based methods learn lightweight detectors
from frequency-domain fingerprints~\cite{gasenzer2023towards} or extract
residual fingerprints from differences between audio and its filtered
version for detection and model attribution~\cite{pizarro2024lightweight}.

\begin{comment}
Detection methods differ in how they obtain acoustic evidence and
allocate computation between the front end and classifier. Pretrained
speech encoders supply learned representations for task-specific
adaptation~\cite{tak2022automatic,kawa2023whisper}, while compact
architectures offer alternatives for speech-specific detection.
AASIST-L models spectral and temporal artifacts with a lightweight
graph-attention network~\cite{jung2022aasist}. Handcrafted cepstral features such as MFCCs~\cite{schafer2025mfcclfcc},
LFCCs~\cite{sahidullah2015features}, and CQCCs~\cite{todisco2016cqcc}
describe short-time spectral characteristics for spoofing detection.
Focusing on vocoder artifacts, Wang and Yamagishi~\cite{wang2023spoofed}
train detectors on speech resynthesized by neural vocoders, using paired
natural and resynthesized recordings for contrastive training.
Gasenzer and Wolter~\cite{gasenzer2024generalizing}
use frequency-domain fingerprints with lightweight detectors, and
Pizarro et al.~\cite{pizarro2025residual} derive residual fingerprints
from differences between signals and their filtered versions for
synthetic-speech detection and attribution. Temporal modulation features
also characterize variations beyond individual short-time
frames~\cite{wu2013modulation,gao2021artifacts}. These studies motivate
combining explicit artifact measurements with complementary acoustic
descriptors.
\end{comment}

\vspace{-3mm}
\section{Artifact-Guided Detection}
\label{sec:method}
\vspace{-1mm}
Our detector combines decoder-guided spectral evidence with complementary
spectro-temporal descriptors. We first analyze decoder upsampling and iSTFT
synthesis to identify frequencies at which generation artifacts may occur
and quantify their evidence in the final waveform. We then characterize
short-time spectral shape and its temporal variation using cepstral features.
The resulting feature groups are fused using a compact gradient-boosted tree.
%We extract complementary spectral and temporal measurements from the final waveform. We analyze transposed-convolution upsampling and iSTFT synthesis to predict where spectral peaks may occur. We measure the strengths of peaks at these frequencies relative to their local background, alongside spectral residuals throughout the analysis band. A shared linear-frequency cepstral front end supplies short-time spectral-shape statistics and Fourier magnitudes of the coefficient trajectories, complementing the time-averaged peak measurements. We concatenate these four feature blocks for a gradient-boosted tree classifier.

\vspace{-3mm}
\subsection{Decoder-Guided Spectral Cues}
\label{sec:Decoder-Induced}
\vspace{-1mm}
\subsubsection{Artifact Origins and Frequencies}
\label{sec:artifact-origins}
\textbf{Decoder Mechanisms.}
Many speech generators synthesize waveforms from low-rate acoustic
representations, such as mel spectrograms or codec embeddings. Learned
upsampling increases temporal resolution, while filters and nonlinear
activations shape the signal. These operations can also introduce periodic
components whose frequencies depend on decoder sampling rates. Some decoders
predict waveform samples directly, as in HiFi-GAN~\cite{kong2020hifi},
whereas others predict spectral coefficients and reconstruct the waveform
using an inverse short-time Fourier transform (iSTFT). We examine how these
operations can produce structured spectral peaks in the output waveform.
Biases and nonlinear activations, such as Leaky ReLU and Snake, can introduce
nonzero temporal means in decoder features. Such a mean corresponds to a
zero-frequency, or DC, component that subsequent upsampling can convert into
tonal artifacts~\cite{pons2021upsampling}.

\noindent\textbf{Transposed-Convolution Artifacts.}
A transposed convolution with stride $s$ is equivalent to inserting $s-1$
zeros between input samples and convolving the expanded sequence with a
learned kernel $h$. For a single input--output channel pair, the response to
a constant input component of amplitude $\mu$, away from boundaries and before
adding the output bias, is $y_{\mu}[sm+r]=\mu G_r$, where
$G_r=\sum_q h[r+qs]$ for $0\leq r<s$, with the sum taken over valid kernel
indices. Because different output phases receive different subsets of kernel
weights, unequal phase sums produce a period-$s$ response when $\mu\neq0$.
This is the one-dimensional checkerboard mechanism expressed through the DC
gains of the polyphase filters~\cite{sugawara2019checkerboard}.

In the frequency domain, zero insertion creates spectral replicas separated
by the input sampling rate $f_{\mathrm{in}}$, while convolution scales them
according to the learned kernel response. A period-$s$ component at output
sampling rate $f_{\mathrm{out}}=s f_{\mathrm{in}}$ can therefore contain
components at $f_k=kf_{\mathrm{in}}$ for positive integers $k$ within the
output Nyquist band. Their amplitudes depend on the feature means, learned
filters, and contributions across channels. Later upsampling stages can
further replicate or attenuate these components~\cite{pons2021upsampling}.

\begin{figure}[!t]
\centering
\includegraphics[width=0.96\linewidth,keepaspectratio]{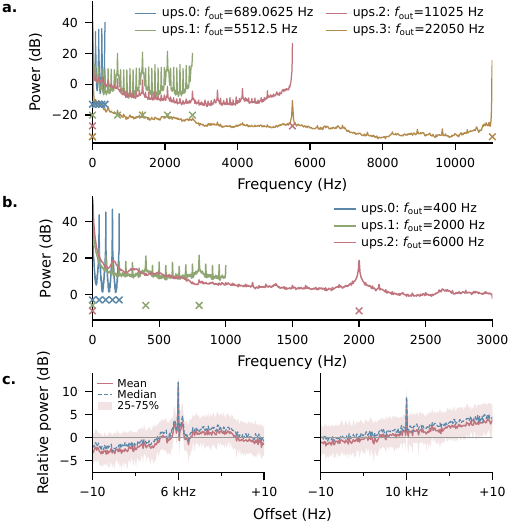}
\vspace{-4mm}
\caption{Upsampling-layer spectra of (a) the OpenVoice V2 tone-color converter and (b) the CosyVoice2 HiFT decoder;
crosses mark predicted DC-replica frequencies. (c) Spectral peaks relative
to the local background at 6 and 10 kHz across 200 CosyVoice2 recordings.}
\vspace{-3mm}
\label{fig:candidate-spectra}
\vspace{-2mm}
\end{figure}

For example, OpenVoice uses a waveform decoder with output sampling rate
$F_s=22050$~Hz and upsampling strides $(8,8,2,2)$~\cite{qin2023openvoice}.
For $L$ layers with strides $s_1,\ldots,s_L$, the input sampling rate of
layer $j$ is $R_{j-1}=F_s/\prod_{\ell=j}^{L}s_{\ell}$. The third OpenVoice
upsampling layer therefore receives features at 5512.5~Hz, giving a possible
DC-replica frequency at 5512.5~Hz. Figure~\ref{fig:candidate-spectra}(a)
shows the spectra of the four upsampling-layer outputs from a single forward
pass.

\noindent\textbf{iSTFT Artifacts.}
For decoders that predict spectral coefficients, artifact locations also
depend on waveform reconstruction. In CosyVoice2, the final upsampling layer
produces features at a sampling rate of 6000~Hz and exhibits a peak at
2000~Hz in Fig.~\ref{fig:candidate-spectra}(b). The decoder predicts
magnitude and phase to form complex spectral coefficients and reconstructs
24-kHz audio using a Hann-windowed iSTFT with $N=16$ and hop
$H=4$~\cite{du2024cosyvoice}.

The spectral coefficients can be decomposed into their temporal mean and a
zero-mean remainder. The mean repeats the same synthesis frame every $H$
samples; away from boundaries, normalized overlap-add therefore produces an
$H$-periodic contribution with possible spectral components at integer multiples of $F_s/H=6000$~Hz.
Periodic variation retained in the zero-mean component can produce additional
replicas around multiples of $F_s/H$. For example, if the 2000-Hz variation
persists in the coefficients, one possible component occurs at
$2(F_s/H)-2000=10000$~Hz.

\noindent\textbf{Waveform Evidence.}
These predicted locations are also visible in generated audio.
Figure~\ref{fig:candidate-spectra}(c) shows narrow spectral peaks near both
6 and 10~kHz in the mean and median spectra of a separate collection of
200 CosyVoice2 recordings, consistent with the decoder analysis. These
waveform measurements use a 10-s rectangular-window FFT at 24~kHz
(0.1-Hz bins), with the waveform mean removed and median log power at offsets
$4<|\Delta f|<40$~Hz subtracted.

\vspace{-3mm}
\subsubsection{Waveform Measurements}
\label{sec:spectral-residuals}
\vspace{-1mm}
We measure decoder artifacts from two complementary perspectives:
a broadband representation of persistent spectral peaks and decoder-guided
measurements at predicted artifact frequencies.

\begin{figure}[!t]
\centering
\includegraphics[width=0.96\linewidth,keepaspectratio]{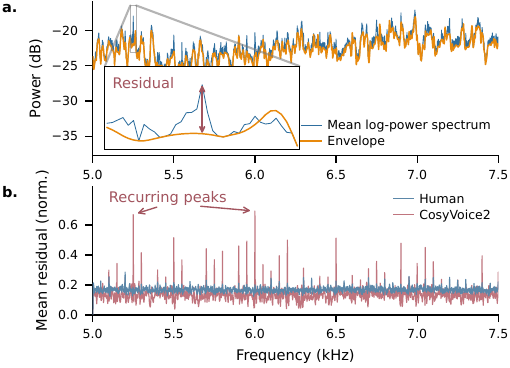}
\vspace{-4mm}
\caption{(a) Background subtraction at 5.25 kHz in a CosyVoice2 recording;
the envelope is the background curve estimated from local spectral minima.
(b) Mean normalized spectral residuals from 200 LibriSeVoc (Human) and 200
CosyVoice2 (TTS) recordings.}
\label{fig:residual-explained}
\vspace{-4mm}
\end{figure}

\noindent\textbf{Broadband Spectral Residuals.}
To capture persistent peaks throughout the analysis band without assuming
their exact locations, we measure their prominence above the surrounding
spectral background. For each 16-kHz input, we compute a Hann-windowed STFT
and average the log-power spectra across frames. Over 1--7.95~kHz, we apply
the background-estimation and normalization procedure of
Afchar et al.~\cite{afchar2025fourier}: a background estimated from local
spectral minima is subtracted, and positive residuals are clipped and
normalized per recording. We retain the normalized residuals at their
original frequency bins, preserving both the locations and relative strengths
of persistent peaks. Figure~\ref{fig:residual-explained}(a)--(b) illustrates
the background subtraction and resulting residual structure.

\noindent\textbf{Decoder-Guided Frequencies.}
We next derive candidate artifact locations from a range of neural vocoder
and codec decoder configurations, including HiFi-GAN~\cite{kong2020hifi},
UnivNet~\cite{jang2021univnet}, and EnCodec~\cite{defossez2022high}.
Applying the relations in Section~\ref{sec:artifact-origins} and merging
locations that fall into the same FFT bin yields a fixed set of 23 candidate
frequencies within 1--7.95~kHz. Their actual amplitudes depend on the learned
filters and subsequent processing, so these frequencies specify where
evidence is measured rather than assuming that every generator exhibits a
strong peak.

\noindent\textbf{Local Peak Prominence.}
At each predicted frequency, we measure peak prominence from the
mean log-power spectrum before background subtraction, clipping, or
normalization. Following the local-background comparison in
Fig.~\ref{fig:candidate-spectra}(c), each frequency is mapped to its
nearest FFT bin. Peak prominence is defined as the maximum log power across
that bin and its two neighbors minus the median log power of background bins
more than 4~Hz and less than 40~Hz away. The neighborhood accommodates
FFT-grid mismatch, while the median estimates the local background.

We additionally compare each candidate peak with nearby spectral
fluctuations. For each predicted location, we subtract the median prominence
at nearby reference centers from the prominence at the candidate frequency.
We use six reference centers on each side, 8--40~Hz from the mapped
frequency. Background bins and reference centers exclude locations within
three FFT bins of registered analysis frequencies and selected replica-grid
positions. We retain both measurements separately for every candidate
frequency, complementing the broadband spectral residuals with
location-specific decoder-guided evidence.

\vspace{-3mm}
\subsection{Spectro-Temporal Features}
\vspace{-2mm}
The decoder-guided measurements above emphasize persistent spectral peaks and
are largely based on temporal averaging. Because such artifacts can vary in
strength across generators, we complement them with short-time spectral shape
and its temporal dynamics. Using 20-ms Hann windows with a 10-ms hop, we
compute log energies from 70 linearly spaced triangular filters over
0--8~kHz and apply a discrete cosine transform, retaining the first
40 linear-frequency cepstral coefficients (LFCCs)
~\cite{sahidullah2015comparison}. Each coefficient forms a temporal
trajectory used by two complementary feature branches.

\noindent\textbf{Distributional Statistics.}
The first branch summarizes the static LFCC trajectories and their first- and
second-order temporal differences. We summarize each trajectory using its median, 10th and 90th
percentiles, and the interdecile range (the 90th percentile minus the 10th percentile), capturing its typical
level and spread while reducing sensitivity to extreme values. These statistics characterize spectral-shape
distributions together with local temporal variation.

\noindent\textbf{Temporal Modulation.}
The second branch characterizes how rapidly spectral shape changes over time,
following modulation-based synthetic-speech
detection~\cite{wu2013synthetic}. For each of the 40 static LFCC trajectories,
we subtract the temporal mean and compute its Fourier magnitude spectrum along
the frame axis. With a frame rate of 100~Hz, we average magnitudes within
eight logarithmically spaced bands spanning 1--50~Hz and apply logarithmic
compression. These modulation features capture the strength of spectral-shape
fluctuations at different temporal rates and complement the distributional
statistics above.

\vspace{-3mm}
\subsection{Feature Fusion \& Classification}
\label{sec}
\vspace{-2mm}
We concatenate four feature groups: broadband spectral residuals,
decoder-guided peak-prominence measurements, LFCC distributional statistics,
and temporal modulation features. The combined representation is classified
using a histogram-based gradient-boosted tree with 1000 boosting iterations,
at most 15 leaves per tree, a learning rate of 0.1, and at least 20 samples
per leaf. Training uses class- and source-balanced weights. We compare feature
configurations on validation data and reserve the test set for final
evaluation.

\vspace{-3mm}
\section{Experiments}
\label{sec:experiment}
\vspace{-2mm}
\subsection{Evaluation Protocols} 
\vspace{-1mm}
We evaluate the detector under three complementary settings. The \textit{seen-source} evaluation measures discrimination when training and test recordings come from the same speech sources but share no recordings. The \textit{cross-source} evaluation tests transfer to previously unseen human and synthetic-speech sources without retraining. Finally, leave-one-generator-out (LOGO) evaluation isolates generator dependence by withholding each synthetic-speech generator from training in turn.

\noindent\textbf{Seen-Source Evaluation.} 
Synthetic speech comes from seven SpeechFake BD subsets ~\cite{huang2025speechfake}: Tortoise, MeloTTS, CosyVoice, OpenVoice, Fish-Speech, StyleTTS2, and ParlerTTS. Human speech comes from the bona fide subset of ASVspoof 5~\cite{wang2024asvspoof} and the ground-truth subset of LibriSeVoc~\cite{sun2023ai}. We split 100,000 development recordings into 80,002 for training and 19,998 for validation, each approximately class-balanced. For our detector, the validation set is further divided into 10,000 tuning and 9,998 calibration recordings. The held-out test set contains 86,234 recordings, equally divided between human and synthetic speech. The training, validation, and test sets share no recordings. Human recordings are drawn 80\% from ASVspoof 5 and 20\% from LibriSeVoc. 

We compare against four compact waveform detectors reported in~\cite{dowerah2026speech}. All four use the same data splits and are trained for up to 100 epochs, with early stopping after 10 epochs without improvement in validation EER. Sampling assigns equal probability to human and synthetic speech and balances sources within each class.

\noindent\textbf{Cross-Source Transfer.} We evaluate all five detectors without retraining and keep their decision thresholds fixed before evaluating the unseen sources. Human speech consists of 1,000 English read-speech recordings from LibriSpeech~\cite{panayotov2015librispeech}. Synthetic speech contains 200 recordings generated by each of four additional TTS systems listed in Table~\ref{tab:models}. The LibriSpeech recordings do not overlap with the LibriSeVoc recordings used for training or seen-source evaluation. 

\noindent\textbf{Leave-One-Generator-Out.} To examine whether the proposed representation depends on generator-specific cues, we retrain the detector in seven folds, each excluding one SpeechFake BD generator from training. Human training data, feature extraction, and classifier hyperparameters remain unchanged. Each fold uses 1000 boosting iterations without validation-based checkpoint selection. We compute EER on the excluded generator together with the same 43,117 human test recordings for every fold (Table~\ref{tab:logo}). 

\noindent\textbf{Preprocessing.} All evaluations use 16-kHz mono audio and the leading 4-s segment of each recording. Only recordings long enough to provide a complete input segment are included. We apply no augmentation, loudness normalization, silence trimming, or additional lossy encoding.

\begin{table}[t]
\centering
\definecolor{eerbg}{HTML}{DCE7F3}
\definecolor{oodbg}{HTML}{DCEEE4}
\definecolor{eerhue}{HTML}{4E86C6}
\definecolor{oodhue}{HTML}{4FA97A}
\caption{Model storage and detection performance:
seen-source EER (\%) and accuracy (\%) at
frozen seen-source thresholds on unseen human speech and four additional
TTS systems.}
\label{tab:models}
\setlength{\tabcolsep}{2pt}
\scriptsize
\newcolumntype{E}{>{\columncolor{eerbg}[2.1pt][2.1pt]}c}
\newcolumntype{O}{>{\columncolor{oodbg}[2.1pt][2.1pt]\centering\arraybackslash}X}
\begin{tabularx}{\columnwidth}{lcEOOOOO}
\toprule
 & & \multicolumn{1}{>{\columncolor{eerbg}[2.1pt][2.1pt]}c}{Seen-Source}
 & \multicolumn{5}{>{\columncolor{oodbg}[2.1pt][2.1pt]}c}{Cross-Source}\\
\cmidrule{3-3}\cmidrule{4-8}
System & Size & EER & Libri & Chatt. & Cosy2 & Qwen3 & Vibe\\
 & & & \cite{panayotov2015librispeech} & \cite{chatterboxtts2025} & \cite{du2024cosyvoice} & \cite{hu2026qwen3} & \cite{peng2025vibevoice}\\
\midrule
RawGAT-ST~\cite{tak2021end} & 1.68\,MiB & \cellcolor{eerhue!52} 0.023 & \cellcolor{oodhue!17} 86.1 & \cellcolor{oodhue!55} 100.0 & \cellcolor{oodhue!55} 100.0 & \cellcolor{oodhue!55} 100.0 & \cellcolor{oodhue!55} 100.0\\
AASIST~\cite{jung2022aasist} & 1.14\,MiB & \cellcolor{eerhue!55} 0.021 & \cellcolor{oodhue!20} 90.7 & \cellcolor{oodhue!55} 100.0 & \cellcolor{oodhue!55} 100.0 & \cellcolor{oodhue!55} 100.0 & \cellcolor{oodhue!55} 100.0\\
RawTFNet~\cite{xiao2025rawtfnet} & 694\,KiB & \cellcolor{eerhue!25} 0.046 & \cellcolor{oodhue!12} 71.8 & \cellcolor{oodhue!55} 100.0 & \cellcolor{oodhue!55} 100.0 & \cellcolor{oodhue!55} 100.0 & \cellcolor{oodhue!55} 100.0\\
W-MesoNet~\cite{kawa2023improved} & 29.2\,MiB & \cellcolor{eerhue!12} 0.065 & \cellcolor{oodhue!16} 84.8 & \cellcolor{oodhue!26} 96.5 & \cellcolor{oodhue!27} 97.0 & \cellcolor{oodhue!22} 94.0 & \cellcolor{oodhue!55} 100.0\\
Proposed & 151\,KiB & \cellcolor{eerhue!55} 0.021 & \cellcolor{oodhue!31} 98.4 & \cellcolor{oodhue!18} 89.0 & \cellcolor{oodhue!39} 99.5 & \cellcolor{oodhue!15} 82.5 & \cellcolor{oodhue!19} 90.0\\
\bottomrule
\end{tabularx}
\vspace{-5mm}
\end{table}

\vspace{-3mm}
\subsection{Detection Performance and Generalization} 
\vspace{-1mm}
\label{sec:results} 
\begin{table}[t]
\centering
\caption{Leave-one-generator-out EER (\%) for proposed detector.}
\label{tab:logo}
\setlength{\tabcolsep}{5pt}
\scriptsize
% Same light blue as the EER block of Table~\ref{tab:models}; needs colortbl.
\definecolor{logohdr}{HTML}{DCE7F3}
\begin{tabular}{lrrrrrrr}
\toprule
\rowcolor{logohdr}
Held out & CosyV. & Tort. & Fish & Parler & StyleT2 & OpenV. & Melo\\
\midrule
Test clips & 8,309 & 10,388 & 2,853 & 1,944 & 2,024 & 7,211 & 10,388\\
EER (\%) & 0.014 & 0.288 & 1.046 & 3.245 & 3.845 & 5.065 & 6.747\\
\bottomrule
\end{tabular}
\vspace{-5mm}
\end{table}

\noindent\textbf{Seen-Source Performance.} The proposed detector achieves an EER of 0.021\%, matching AASIST, the best-performing neural comparator in this evaluation, while requiring an estimated 151~KiB of tree storage compared with 694~KiB--29.2~MiB for the neural models (Table~\ref{tab:models}). This result shows that explicit acoustic measurements can provide highly discriminative evidence on seen speech sources without learning a waveform feature extractor end to end. 

\noindent\textbf{Cross-Source Transfer.} Performance becomes more source dependent when the trained models are applied without retraining to unseen speech sources. The proposed detector correctly accepts 98.4\% of the unseen LibriSpeech recordings as human, compared with 71.8--90.7\% for the neural comparators. Its recall on unseen synthetic speech, however, varies from 82.5\% on Qwen3-TTS to 99.5\% on CosyVoice2, whereas the neural comparators reach 94--100\% across the four synthetic sources. The resulting asymmetry is informative: the explicit representation produces relatively few false alarms on unseen human speech, but transfer to unseen generators is less uniform. Thus, generator shift, rather than seen-source discrimination, is the more important limitation of the current screen. 

\noindent\textbf{Held-Out Generator Generalization.} The LOGO experiment further isolates this effect. EER ranges from 0.014\% when CosyVoice is excluded from training to 6.747\% when MeloTTS is excluded, with an average of 2.893\% (Table~\ref{tab:logo}). The wide range indicates that cues learned from the remaining generators transfer strongly to some held-out systems but not to others. This is consistent with the decoder analysis in Section~\ref{sec:artifact-origins}: related synthesis operations can create shared spectral structure, while learned filters, intermediate representations, and waveform reconstruction can change the strength and shape of the artifacts. These results motivate evaluating the screen at explicit miss-rate constraints rather than relying on EER alone.

\vspace{-3mm}
\subsection{Screening Efficiency}
\label{sec:energy}
\vspace{-1mm}
Because the proposed model is designed as a front-end screen, its practical
benefit depends on how many recordings can bypass expensive backend inference
while maintaining a low synthetic-speech miss rate. We therefore evaluate a
two-stage cascade in which our model screens every request and forwards
suspected synthetic speech to \textit{DF\_Arena\_1B\_V\_1} (DF), a
1.15-billion-parameter backend detector~\cite{dowerah2026speech}. For a
workload with synthetic-speech prevalence $p$, the fraction forwarded to DF is
$q=(1-p)\mathrm{FPR}+p(1-\mathrm{FNR})$. Using a reference workload with
$p=0.1$, the 0.050\% miss-rate operating point in
Table~\ref{tab:screening-points} gives $q\approx0.100035$, allowing about
90\% of requests to bypass the backend.

We measure DF inference energy on an NVIDIA A100-SXM4-80GB using 256
randomly selected, class-balanced 4-s recordings at batch size 1. Across
three post-warmup measurement windows, integrating GPU board power above
the preceding idle baseline gives 10.422~J/request, including input transfer
and score return. This subset is used only to estimate backend inference cost;
forwarding rates are computed from the full test set. Front-end CPU costs are
estimated from single-core AMD Milan runtimes assuming 15~W: audio decoding
requires 0.310~J/request in both workflows, while feature extraction and tree
classification add 0.324~J/request.

Let $E_{\mathrm{decode}}$, $E_{\mathrm{our}}$, and $E_{\mathrm{DF}}$ denote
the corresponding per-request costs. Direct backend detection requires
$E_{\mathrm{direct}}=N(E_{\mathrm{decode}}+E_{\mathrm{DF}})$, whereas the
two-stage cascade requires
$E_{\mathrm{screen}}=N(E_{\mathrm{decode}}+E_{\mathrm{our}}
+qE_{\mathrm{DF}})$. For $N=10^6$ requests, estimated energy decreases from
2.981 to 0.466~kWh, an 84.4\% reduction. The reduction primarily results from
avoiding backend inference on benign requests while maintaining a stringent
synthetic-speech miss rate. This estimate assumes constant DF energy per request and excludes
model loading, warmup, GPU idle energy, and DF host-side costs.

% Generated by icassp_row/export_screening_table.py; rates are percentages.
\begin{table}[t]
\centering
\caption{Screening operating points selected from the held-out test ROC
under the specified empirical FNR limits.}
\label{tab:screening-points}
\setlength{\tabcolsep}{6pt}
\scriptsize
% Same light blue as the EER block of Table~\ref{tab:models}; needs colortbl.
\definecolor{scrcol}{HTML}{DCE7F3}
\begin{tabular}{>{\columncolor{scrcol}}lrrrr}
\toprule
FNR limit (\%) & 0.000 & 0.010 & 0.020 & 0.050\\
\midrule
FPR (\%) & 0.359 & 0.058 & 0.021 & 0.009\\
\bottomrule
\end{tabular}
\vspace{-5mm}
\end{table}

\vspace{-3mm}
\section{Conclusion}
\vspace{-1mm}

In this paper, we presented a compact synthetic-speech screening approach that combines
decoder-guided spectral artifacts with complementary spectro-temporal
features. The resulting detector achieves 0.021\% seen-source EER with an
estimated model size of 151~KiB, while a two-stage screening analysis shows
the potential to substantially reduce expensive backend inference. Results
on unseen generators further highlight that cross-generator transfer remains
the main challenge, motivating future work on more robust artifact
representations and adaptive screening across evolving synthesis systems.
\newpage
\bibliographystyle{IEEEbib}
\bibliography{references}

\clearpage
% \appendix
% \input{appendix_feature_details-v2.tex}
\end{document}